\documentclass[aps,prl,twocolumn,floats,showpacs,superscriptaddress]{revtex4}
\usepackage{graphicx,epsfig}% Include figure files
\usepackage{times}
\usepackage{graphics,dcolumn,bm,float}
\usepackage{amssymb,amsmath,rotate,color}

\begin{document}
\unitlength 1 cm
\newcommand{\be}{\begin{equation}}
\newcommand{\ee}{\end{equation}}
\newcommand{\bearr}{\begin{eqnarray}}
\newcommand{\eearr}{\end{eqnarray}}
\newcommand{\nn}{\nonumber}
\newcommand{\vk}{\vec k}
\newcommand{\vp}{\vec p}
\newcommand{\vq}{\vec q}
\newcommand{\vkp}{\vec {k'}}
\newcommand{\vpp}{\vec {p'}}
\newcommand{\vqp}{\vec {q'}}
\newcommand{\bk}{{\bf k}}
\newcommand{\bp}{{\bf p}}
\newcommand{\bq}{{\bf q}}
\newcommand{\br}{{\bf r}}
\newcommand{\up}{\uparrow}
\newcommand{\down}{\downarrow}
\newcommand{\fns}{\footnotesize}
\newcommand{\ns}{\normalsize}
\newcommand{\cdag}{c^{\dagger}}

\definecolor{red}{rgb}{1.0,0.0,0.0}
\definecolor{green}{rgb}{0.0,1.0,0.0}
\definecolor{blue}{rgb}{0.0,0.0,1.0}

\title{Anderson Transition in Disordered Graphene}

\author{Mohsen Amini}
%\footnote {Electronic address:
%msn\_amini@ph.iut.ac.ir}}}

\affiliation{Department of Physics, Isfahan University of
Technology, Isfahan 84154-83111, Iran}

\author{S. A. Jafari}
%{\footnote {Electronic address:
%sa.jafari@cc.iut.ac.ir}}}

\affiliation{Department of Physics, Isfahan University of
Technology, Isfahan 84154-83111, Iran} \affiliation{The Abdus
Salam ICTP, 34100 Trieste, Italy}

\author{Farhad Shahbazi}
%{\footnote {Electronic address:
%shahbazi@cc.iut.ac.ir}}}

\affiliation{Department of Physics, Isfahan University of
Technology, Isfahan 84154-83111, Iran}

%\maketitle   %revtex
\pacs{ 72.15.Rn, 72.20.Ee, 81.05.Uw }
\begin{abstract}
We use the regularized kernel polynomial method (RKPM) to
numerically study the effect disorder on a single layer of
graphene. This accurate numerical method enables us to study very
large lattices with millions of sites, and hence is
almost free of finite size errors. Within this approach, both
weak and strong disorder regimes are handled on the same footing.
We study the tight-binding  model with on-site disorder, on the
honeycomb lattice. We find that in the weak disorder regime, the
Dirac fermions remain extended and their velocities decrease as the disorder
strength is increased. However,  if the disorder is strong enough, there will
be a {\em mobility edge} separating  {\em localized states around
the Fermi point}, from the remaining extended states. This is in
contrast to the scaling theory of localization which predicts
that all states are localized  in two-dimensions (2D). 

%Our results suggest
%the possibility of making  semiconductor from single-layer
%graphene, with intrinsic acceptor and donor levels, by simply
%introducing a proper amount of disorder to the sample.
\end{abstract}

\maketitle  %revtex4
\pagebreak

{\em Introduction}:Graphene  is the 2D allotrope  of
 carbon  in which carbon atoms with $sp^{2}$ hybridization are organized in a
honeycomb lattice. Recently, the isolation of single graphene
sheets  has become possible through chemical exfoliation of
bulk graphite~\cite{Novoselov1,Novoselov2,Geim}, or  epitaxial
growth, either by chemical vapor deposition of hydrocarbons  on
metal substrates~\cite{cvd} or by  thermal decomposition of
SiC~\cite{Berger}.  The latter method produces graphene layers with high
crystalline quality.  Properties such as, tuning the carrier
types  from hole to electron in the same sample through a gate
voltage and remarkably high mobility of charge carriers, even at
room temperature, due to the absence of back-scatterings, has made
the graphene promissing from the technological point of
view in building the carbon-based  electronic
devices~\cite{NetoRMP}. 

%Graphene ribbons have also similar
%properties to those of carbon nano-tubes, that is they exhibit
%metallic, as well as semi-conducting behavior depending on their
%shapes~\cite{Xu}. However, converting the infinite single-layer
%graphene to a semiconductor still remains as an open
%problem~\cite{Lanzara,NetoPRB2008}.

  The electronic properties of pure graphene can be modeled 
with a simple tight-binding picture proposed by
Wallace~\cite{Wallace} which provides the essential band
structure of graphene as a half-filled system with a density of
states (DOS) vanishing linearly at the charge neutrality point.
In this picture the energy dispersion is linear in momentum near
the Fermi points, causing the quasi-particles behave as
mass-less chiral Dirac electrons~\cite{semenof}. 

%There is an alternative
%approach based on the concept of resonating valance bond  (RVB)
%concept proposed by Pauling~\cite{Pauling} and advocated by
%others~\cite{BaskaranJafari}, which provides some insight into the
%nature of many-body excitations in graphene in strong coupling
%regime.

 In spite of high crystalline quality, disorder can not be avoided
in currently available samples of graphene. Various kinds of
intrinsic disorder such as surface ripples and topological
defects, as well as extrinsic ones like  vacancies, ad-atoms,
charge impurities
affect the electronic properties of graphene.  The presence of a
minimal conductivity at zero bias~\cite{Novoselov2} implies the
existence of extended states at the charge neutrality points,
in contrast to prediction of the scaling theory
in 2D~\cite{AndersonLocalization}. 

There has been
various theoretical and numerical approaches to study the effect
of different types of disorder in graphene~(as a review,  see
Ref.~\cite{NetoRMP} and references therein). Based on
one-parameter scaling theory~\cite{LeeRama}, Ostrovsky and
coworkers found a metal-insulator transition in
graphene for long range disorder~\cite{Ostrovsky}. The hypothesis of one-parameter
scaling was subsequently verified with numerical calculation of
Bardarson et al.~\cite{Bardarson}. They used a transfer operator
method to investigate the localization properties of the
low-energy effective (Dirac) theory. Similar result for the beta
function was obtained independently by Nomura et
al.~\cite{NomuraRyu} by evaluating the Kubo formula for the
conductance. 
Although these numerical works verify the hypothesis of
single-scaling  theory, but they predict that Dirac fermions
remain delocalized at arbitrary strength of disorder. 
The effect of roughness on the
electronic conductivity was studied in~\cite{Abedpour}, and it
was found that all states remain localized in the presence of
random effective gauge fields induced by ripples.

Lherbier et al. used a real space and order $N$ Kubo formalism to
calculate time dependent diffusion constant $D(E,t)$ for the Anderson model on 
honeycomb lattice~\cite{Roche}. 
For small values of disorder strength $W$, they found $D(E,t)$ saturates to a 
constant value in long time limit, indicating the presence of extended states.
At larger values of $W$ they report a decrease in $D(E,t)$ pinpointing the
onset of localization.

 Most of the analytical methods used in the studies of disorder in graphene, are able to
handle restricted regimes of specific types of disorder. Their
predictions are valid only on the low-energy scales around the
Dirac points, where  the inter-valley scatterings from the
impurity potential can be ignored. When disorder strength becomes
comparable to the band-width, it is important to take into
account all energy scales simultaneously along with possible
interplay between different energy scales. There also remains an
important question that, whether there is a mobility edge in
graphene in the strong disorder regime or not?

 In this paper we use the  kernel polynomial method
(KPM)~\cite{KPM} based on the expansion of spectral functions in
terms of a complete set of polynomials to accurately calculate
various spectral properties, including the density of states
(DOS). The computation time in KPM method grows linearly with
system size. Matrix manipulations can be done on the fly, which
reduces the memory usage enormously. Therefore one can study very
large lattice sizes in a moderate time.  Regularization of KPM
method known as RKPM remedies the Gibbs oscillations~\cite{RKPM},
and therefore is capable to handle {\em any type of disorder}
with {\em arbitrary strength} in essentially exact way. By this
method, one can treat the low-energy and high-energy features of
graphene on the same footing, and hence the interplay between the
Dirac features and high energy parts of the spectrum, as well as
inter-valley scattering is taken into account. This method
enables us to explore new regimes of disorder strength with
fascinating properties.

{\em Model and method:} From the single-particle point of view, disorder can
affect the non-interacting electrons in graphene, mainly  through spatial
variations off on-site energy (diagonal disorder) or changes in the hopping
integrals due to the variations in the distances or angles of the $p_{z}$
orbitals (off-diagonal disorder). In what follows we study the graphene with
on-site uncorrelated disorder. We consider non-interacting electrons moving on
a honeycomb lattice in the presence of local diagonal disorder. The basic model
to describe this kind of problem is the Anderson model:
  \be
   H =-t\sum_{\langle i,j\rangle}[c_{i}^\dagger c_{j} + H.c] + \sum_{j}^{N}
   \epsilon_{j} c_{j}^\dagger c_{j}. \label{hamiltoni}
   \ee
The first term
describes  hopping between nearest-neighbor sites  and $\epsilon_{i}$'s in the
second term are the random on-site potential  uniformly distributed in the
interval $[-W/2,W/2]$. In Eq.~(\ref{hamiltoni}), the energy $t$ is associated
with nearest neighbor hopping integral, which is about $2.7$ eV in
graphene. This model has recently been
studied by transfer-matrix method in Ref.~\cite{Xiong}, where it was found
that all states are localized, in agreement with scaling theory of localization
in 2D. However, our results are in contradiction with this finding.

In our work,  to investigate the localization properties, we employ the so
called {\em typical DOS} as a quantity which determines whether a given state
with energy $E$ is extended or localized~\cite{KPM}. The return probability for
an extended state at a given energy $E$ is zero~\cite{PhilipsBook}. Therefore
the self-energy of an extended state acquires an imaginary part to account for
the decay of return probability, while the self-energy for localized states
remains purely real. This reflects itself in the local density of states
(LDOS):
\be \rho_s(E) = \sum_k |\langle s|E_k\rangle|^2\delta(E-E_k),
\label{LDOS}
\ee
where $|s\rangle$ denotes a localized basis state at site $s$,
while $|E_k\rangle$ is an energy eigen-vector corresponding to energy $E_k$.
Examining LDOS at $K_s$ sites provides a measure to distinguish localized states from
extended ones. To do so, one defines the  typical DOS
which is a geometrical average of LDOS's,
\be \rho_{\rm typ}(E) =
  \exp\left[\frac{1}{K_rK_s}\sum_r^{K_r}\sum_s^{K_s}\ln\left(\rho_s^{r}(E)\right)\right],
   \label{TYPDOS}
\ee
where $K_r$ is the number
of realizations used in numerical calculations. We also need the total spectral
weight which can be obtained from the following arithmetic averaging: \be
\rho(E) = \frac{1}{K_rK_s}\sum_r^{K_r}\sum_s^{K_s}\rho_s^{r}(E)
=\frac{1}{D}\sum_{k=0}^{D-1}\delta(E-E_k), \label{AVDOS} \ee where $D$ is the
dimension of the Hilbert space on which the Hamiltonian $H$ is acting.

The core quantity in both Eqs.~(\ref{TYPDOS}) and~(\ref{AVDOS}) is the LDOS.
In order to calculate quantities of this sort, the KPM method~\cite{KPM}
employs a complete basis set.

The basic idea behind the KPM is to expand the
spectral function, say, $\rho_s(E)$,  in terms of orthogonal polynomials, $\phi_{m}(E)$,
In general, all
types of orthogonal polynomials can be used. In the case
of e.g. Chebyshev polynomials one has:
   \be
   \rho_s(E)=\frac{1}{\pi\sqrt{1-E^2}}\left[\mu_0+2\sum_{m=1}^M \mu_m T_m(E)\right],
   \label{rhos.eqn}
   \ee
where
   \be
   \mu_m=\int_{-1}^1 \rho_s(E) T_m(E)dE
   =\frac{1}{D}\langle s|T_m(\tilde{H})|s\rangle. \label{muLDOS}
   \ee
Here, $\tilde H$ is obtained from $H$ by a simple linear transformation to ensure
that the eigenvalues of $\tilde H$ are in $[-1,1]$.  The same procedure when
applied to Eq.~(\ref{AVDOS}) gives:
\be
   \mu_m = \int_{-1}^1 \rho(E) T_m(E)
   dE =\frac{1}{D} \mbox{Tr}[T_m(\tilde H)].  \label{muDOS}
\ee
Eqs.~(\ref{muDOS}) and (\ref{muLDOS}) can be evaluated with a recursive
relation first discussed by Wang~\cite{Wang}. The Tr in
Eq.~(\ref{muDOS}) can be most conveniently evaluated with a simple stochastic
summation employing the recursion relation among Chebyshev
polynomials~\cite{KPM}.  However, since $M$ in Eq.~(\ref{rhos.eqn}) is finite in
computer implementations, one faces the classic problem of Gibbs oscillations.
There are standard attenuation factors suggested in the literature which
minimize such unwanted oscillations~\cite{Wang, KPM}. Due to peculiar Dirac
dispersion in graphene, none of the so called $g-$factors worked. The solution
around this problem is to use regulated (Gaussian broadened) polynomials to
calculate the moments~\cite{RKPM}: \be \langle \phi_{m}(x)\rangle_{\sigma}
\equiv \frac{1}{2\pi\sigma^2}\int dx' e^{-(x'-x)^2/2\sigma^2}\phi_{m}(x'),
\label{RP} \ee Our present calculation is based on using Legendre polynomials
in (\ref{RP}) with $\sigma$ equal to $4/M$.

\begin{figure}[t] \includegraphics[angle=-90,width=8.5cm]{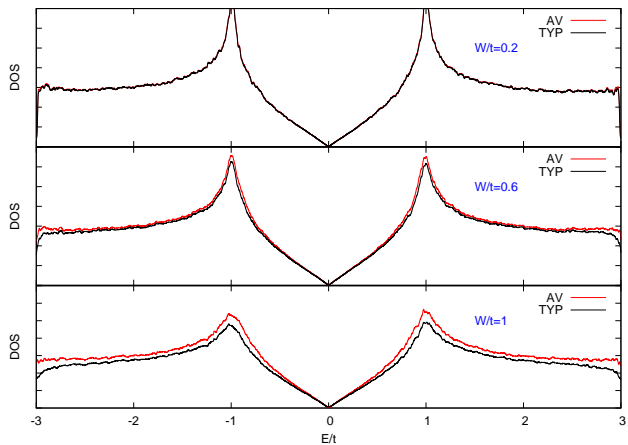}
\caption{(Color online) $\rho$ (solid line) and $\rho_{\rm typ}$
(dotted line) for different values of disorder strength $W$ in
the weak disordered regime ($W \lesssim t$). In this regime the
disorder only renormalizes the Fermi velocity of Dirac
quasi-particles. Our results are stable with respect to change in
both lattice size, as well as $M$. } \label{weak.fig}
\end{figure}

\begin{figure}[t]
\includegraphics[angle=-90,width=8.5cm,clip=true]{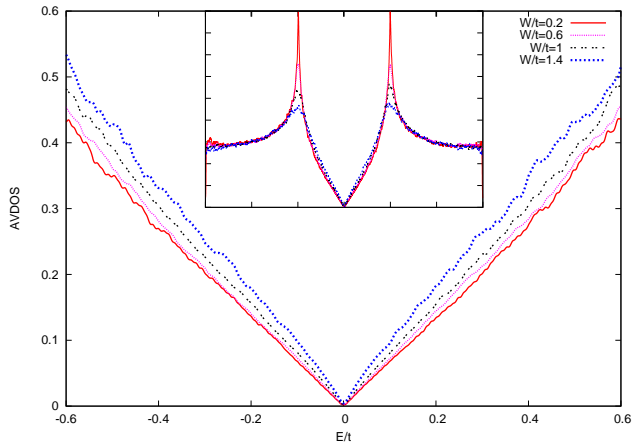}
\caption{(Color online) Average DOS as a function of energy.
By increasing $W$ in weakly disordered region, the velocity of Dirac fermions
decreases.}
\label{dirac.fig}
\end{figure}

{\em Results and discussions}: To investigate  the Anderson transition in
graphene, we use Eq.~(\ref{hamiltoni}) with different values of $W$ and then
calculate  the LDOS via the RKPM method with $M$ equal to $5000-12000$ on a
lattice with $10^6$ sites.  The result for different values of $W$ is shown in
Fig.~\ref{weak.fig}.
%%%%%%%%%%%%%%%%%%%%%%%%%%%%%%%%%%%%%%%%%%%%%%%%%%%%%%%%%%%%%%%%%%%%%%%%%%%%%%%%%%%%%%%%%%%%%%%%
%%%%%%%%%%%%%%%%%%%%%%%%%%%%%%%%%%%%%%%%%%%%%%%%%%%%%%%%%%%%%%%%%%%%%%%%%%%%%%%%%%%%%%%%%%%%%%%%%%
   As can be seen in Fig.~\ref{weak.fig} the average DOS for various values of
$W \lesssim t$ resembles that of perfect graphene. The typical
DOS, $\rho_{\rm typ}$, is non-zero every where, indicating that
none of the states are localized in this regime. The role of
disorder at weakly disordered regime is to slow down the Dirac
quasi-particles, with their Dirac nature preserved. This result
is in agreement with other
works~\cite{DasSarmaBornApprox,Qaiumzadeh}. Fig.~\ref{dirac.fig}
shows the renormalization of the Dirac fermion's velocity. By
increasing the disorder strength, the slope of $\rho$ decreases.
Therefore, according to $\rho(\omega)\propto |\omega|/v_F^2$, the
velocity of quasi-particles decreases when $W$
increases~\cite{DasSarmaBornApprox,Roche}.

%%%%%%%%%%%%%%%%%%%%%%%%%%%%%%%%%%%%%%%%%%%%%%%%%%%%%%%%%%%%%%%%%%%%%%%%%%%%%%%%%%%%%%%
%%%%%%%%%%%%%%%%%%%%%%%%%%%%%%%%%%%%%%%%%%%%%%%%%%%%%%%%%%%%%%%%%%%%%%%%%%%%%%%%%%%%%%%

\begin{figure}[t]
\includegraphics[angle=-90,width=8.cm,clip=true]{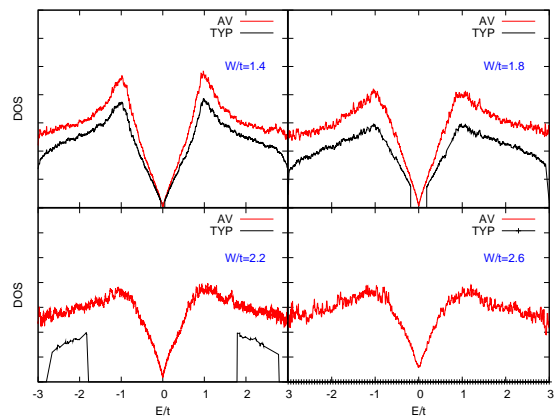}
\caption{(Color online) For $W > t$ a mobility edge starts to
appear. $\rho_{\rm typ}$ is zero around the Fermi energy.
Therefore localized states fall into a "gap" separating extended
states of the valance and conduction bands.} \label{strong.fig}
\end{figure}

   Increasing $W$ to the strong disorder regime, $W \gtrsim t$, we observe a mobility
edge in graphene in Fig.~\ref{strong.fig}.
The mobility edge starts at the Fermi energy and keeps moving to
both left and right side separating localized states around Fermi
energy from those at higher energies. This is unlike the usual
scenario of localization where states at the band edge start to
localize first. It has to do with the geometrical nature of
graphene lattice where states at Fermi energy correspond to
momenta around the $K$ points with their wave-length of the order
of atomic separation. Upper edge of the conduction band and lower
edge of the valance band of clean graphene arise from $\Gamma$
point of the Brillouin zone which correspond to small momenta,
or long wave-lengths, which naturally localize later than short
wave-lengths modes. Such peculiar band picture in graphene leads
to a particle-hole continuum in graphene which is drastically
different from usual continua of Fermi liquids with extended
Fermi surface~\cite{BaskaranJafari}.

%%%%%%%%%%%%%%%%%%%%%%%%%%%%%%%%%%%%%%%%%%%%%%%%%%%%%%%%%%%%%%%%%%%%%%%%%%%%%%%%%%%%%%%%%%%%%%%%%%%%%
%%%%%%%%%%%%%%%%%%%%%%%%%%%%%%%%%%%%%%%%%%%%%%%%%%%%%%%%%%%%%%%%%%%%%%%%%%%%%%%%%%%%%%%%%%%%%%%%%%%%%%

   The idea of band gap opening in graphene
has already been proposed by several groups.
Pereira {\em et al}~\cite{NetoPRB2008} by 
selectively producing vacant sites  in only one sub-lattice,
found the appearance of a  clean gap without any states in it. 
Substrate induced band gap,  which is
accompanied by breaking the particle-hole symmetry is another
possible scenario~\cite{Lanzara}. 

\begin{figure}[t]
\includegraphics[angle=-90,width=8.cm,clip=true]{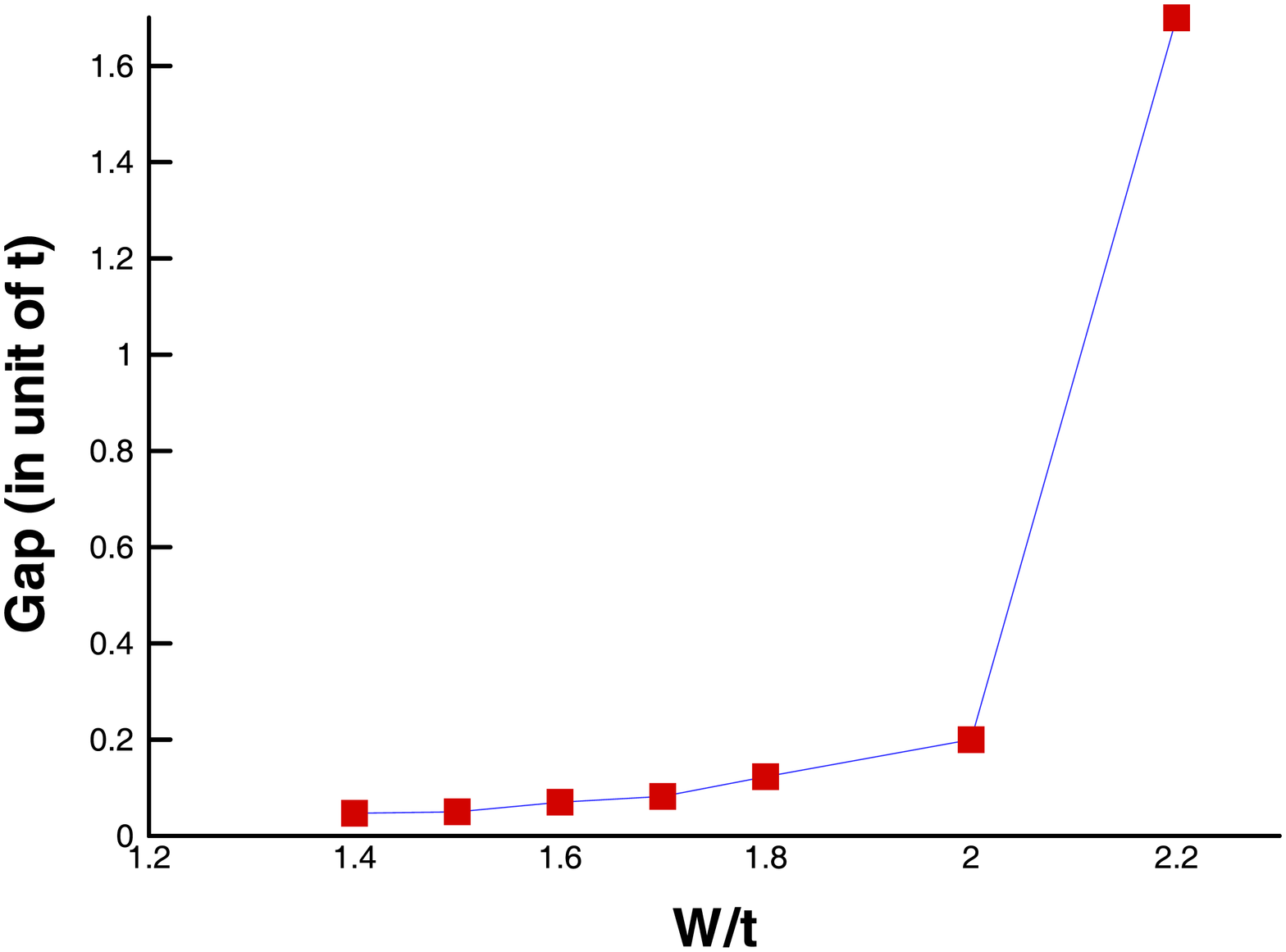}
\caption{(Color online) Gap (distance between the mobility edges
around the neutrality point) versus disorder strength.}
\label{gap.fig}
\end{figure}

However, our result suggest the possibility of {\em disorder induced gap} 
in graphene. This gap is defined as the distance between the upper and 
lower mobility  edges around the Fermi point of graphene.
We find that in the intermediate disorder regime ($W\sim t$)
states at the Fermi point start to localize, thereby opening a
small gap around it (Fig.~\ref{strong.fig}). 
The states inside the gap are localized and
 can not contribute in transport phenomena. Upon increasing
disorder strength, this gap rises slowly up to $W\sim 2t$, above
which its  growth speeds up  with a high slope (Fig.~{\ref{gap.fig}}). This behavior can be assigned to
cross-over from weak to strong localization regimes. 
Naumis also obtains such disorder induced gap, using a real space 
renormalization group scheme ~\cite{Naumis} for vacancy doped graphene.
Nevertheless this gap continuously grows from zero as a function of doping ratio.

Finally with increasing $W$ beyond $W_c/t = 2.5\pm 0.5$, 
all states in the band become localized.
The "error" $\pm 0.5$ here needs clarification: One might argue
that the typical DOS is not the best quantity to distinguish
localized states from extended ones. Our benchmark runs for the
known results of the critical $W$ for the 3D cubic systems gives
$W_c^{\rm 3D}/t=16.0\pm 0.5$. Therefore the possible intrinsic
error in finding $W_c$ in this method is less than $0.5$.

{\em Conclusion}:
   For the Anderson model, in weak disorder regime, we observe that
the Dirac fermions remain delocalized up to $W^* \approx t$. In
this regime, the effect of disorder is to 
decrease the  velocity of Dirac fermions, hence resulting in a
renormalized Dirac cone.
However, upon increasing the disorder strength beyond $W^*$ in
this model, we observe a mobility edge which supports the beta
function proposed in Ref.~\cite{Ostrovsky}. Our results support
the idea proposed by Suzuura and Ando~\cite{Suzuura} which explains
the suppression of weak localization in graphene. According to
their argument, the change in relative weights of the two
components of chiral electrons  wave functions induces a new
Berry phase, when these electrons move along a closed path. In
the absence of inter-valley scatterings (weak disorder regime),
this new phase changes the sign of the wave-function in a given
path with respect to its time reversed counterpart, hence  leading
to a destructive interference of the two paths. This is the
reason for  the existence of  extended states close to the charge
neutrality point in weak disorder regime. We think the
localization of the all electronic states, obtained by transfer
matrix method~\cite{Xiong}, might be due to the missing of the
chirality effect. When the disorder width is comparable with the
hopping energy ($W\sim t$), the inter-valley scatterings are
present and the weak-localization will be eventually recovered
around the Fermi points~\cite{Aleiner,Altland,Louis,Falko}. In very
strong disordered regime ($W > 2t$), localization quickly
spreads  over all the energy spectrum. In this case  the system
is no longer homogeneous in a sense that it divides into regions
with different chemical potential and transport is described in
terms of percolation in the real-space~\cite{Cheianov,Shklovskii}.

{\em Acknowledgements}: This work was partially supported by ALAVI
Group Ltd. We are also indebted to S. Sota, R. Asgari and F.
Fazileh  for useful discussions.

%%%%%%%%%%%%%%%%%%%%%%%%%%%%%%%%%%%%%%%%%%%%%%%%%%%%%%%%%%%%%%%%%%%%%%%%%%%%%%%%%%%%%%%%%%%%%%%%%%%%%
%%%%%%%%%%%%%%%%%%%%%%%%%%%%%%%%%%%%%%%%%%%%%%%%%%%%%%%%%%%%%%%%%%%%%%%%%%%%%%%%%%%%%%%%%%%%%%%%%%%%%%

%\pagebreak


\begin{thebibliography}{50}

\bibitem{Novoselov1} K. S. Novoselov et al., Science  {\bf 306}, 666 (2004).

\bibitem{Novoselov2}  K. S. Novoselov et al., Nature  {\bf 438}, 197 (2005).

\bibitem{Geim}A. K. Geim and K. S. Novoselov, Nature Mater. {\bf 6}, 183-191 (2007).


\bibitem{cvd} T. A. Land, T. Michely, R. J. Behm, C. J. Hemminger, and G. Comsa,  Surf. Sci. {\bf 264}, 261 (1992);
A. Nagashima, {\it et al}, Surf. Sci. {\bf 291}, 93 (1993).

\bibitem{Berger} C. Berger {\it et al}, J. Phys. Chem. B {\bf
108}, 19912 (2004); C. Berger {\it et al}, Science {\bf 312},
1191 (2006)

\bibitem{NetoRMP} A. H. Castro Neto, F. Guinea, N. M. R. Peres, K. S. Novoselov, A. K. Geim,
arXiv:0709.1163v2.

\bibitem{Wallace} P. R. Wallace, Phys. Rev. {\bf 71}, 622 (1947).

\bibitem{semenof} G. W. Semenof, Phys. Rev. Lett. {\bf 53}, 2449 (1984).

\bibitem{AndersonLocalization} E. Abrahams, P. W. Anderson, D. C. Licciardello, and T. V. Ramakrishnan,
Phys. Rev. Lett. {\bf 42}, 673(1979).

\bibitem{LeeRama} P. A. Lee, T. V. Ramakrishnan, Rev. Mod. Phys. {\bf 57} 287 (1985).

\bibitem{Ostrovsky} P. M. Ostrovsky, I. V. Gornyi, A. D. Mirlin, Phys. Rev. Lett. {\bf 98}, 256801 (2007).

\bibitem{Bardarson} J. H. Bardarson, J. Tworzydlo, P. W. Brouwer, C. W. J. Beenakker, Phys. Rev. Lett. {\bf 99}, 106801 (2007);

\bibitem{NomuraRyu} K. Nomura, M. Koshino, S. Ryu, Phys. Rev. Lett. {\bf 99}, 146806 (2007).

\bibitem{Abedpour} N. Abedpour, et al, Phys. Rev. B {\bf 76}, 195407 (2007).

\bibitem{Roche} A. Lherbier, et al. Phy. Rev. Lett. {\bf 100}, 036803 (2008); F. Triozon et al.,
Phys. Rev. B {\bf 69}, 121410 (2004).

\bibitem{KPM} A. Weisse et al., Rev. Mod. Phys. {\bf 78}, 275 (2006).

\bibitem{RKPM} Shigetoshi Sota, Masaki Itoh, JPSJ. {\bf 76}, 054004 (2007).

\bibitem{Xiong} S. J. Xiong, and Y. Xiong, Phys. Rev. B {\bf 76}, 214204 (2007).

\bibitem{PhilipsBook} P. Philips, {\em Advanced Solid State Physics},
Westview Press, 1st Ed. (2002).

\bibitem{Wang} Lin-Wang Wang, Phys. Rev. B {\bf 49}, 10154 (1994).

\bibitem{DasSarmaBornApprox} B. Yu-Kuang Hu, E. H. Hwang, S. Das Sarma, arXiv:0805.2148.

\bibitem{Qaiumzadeh} A. Qaiumzadeh, N. Arabchi, R. Asgari, arXiv:0805.3890

\bibitem{BaskaranJafari} G. Baskaran, S. A. Jafari, Phys. Rev. Lett. {\bf 89} 016402 (2002);
 {\bf 92} 199702 (2004); N.M.R. Peres, et al., {\em ibid}, {\bf 92} 199701 (2004).

\bibitem{NetoPRB2008} Vitor M. Pereira, J. M. B. Lopes dos Santos, A. H. Castro Neto
Phys. Rev. B {\bf 77} 115109 (2008).

\bibitem{Lanzara} S. Y. Zhou, {\em et al}, Nature Materials {\bf 6}, 770-775 (2007);
{\em ibid} {\em 7}, 259-260 (2008).

\bibitem{Naumis} G. G. Naumis, Phys. Rev. B {\bf 76}, 153403 (2007).

\bibitem{Suzuura} H. Suzuura, and T. Ando, Phys. Rev. Lett. {\bf 89}, 266603 (2002).

\bibitem{Altland} A. Altland, Phys. Rev. Lett. {\bf 97}, 236802 (2006).

\bibitem{Aleiner} I. L. Aleiner, and K. B. Efetov, Phys. Rev.
Lett. {\bf 97}, 236801 (2006).

\bibitem{Louis} E. J. Louis, A. Verg\'es, F. Guinea, and G.
Chiappe, Phys. Rev. B {\bf 75}, 085440 (2007).

\bibitem{Falko} E. McCann, et al., Phys. Rev. Lett. {\bf 97}, 146805 (2006);
K. Kechedzhi, et al.  Eur. Phys. J. Special Topics {\bf 148}, 39–54 (2007).

\bibitem{Cheianov} V. V. Cheianov, V. I. Fal'ko, B. L. Altshuler,
and I. L. Aleiner, Phys. Rev. Lett. {\bf 99}, 176801 (2007).

\bibitem{Shklovskii} B. I. Shklovskii, eprint arXive:0706.4425
(2007).

\end{thebibliography}
\end{document}